\documentstyle[prb,aps,epsf,floats,goodfloat]{revtex}
\advance\topmargin by 0.8in
\newcommand{\smfrac}[2]{\mbox{\small $#1 \over #2$}}
\def\xit{\xi_{\rm typ}}
\def\xmin{x_{\rm min}}
\def\e0{\epsilon_0}
\begin{document}
\draft

\twocolumn[\hsize\textwidth\columnwidth\hsize\csname@twocolumnfalse%
\endcsname

\title{
Finite Temperature and Dynamical
Properties of the Random Transverse-Field Ising Spin Chain}

\author{A. P. Young}
\address{Department of Physics, University of California, Santa Cruz, 
CA 95064}

\date{\today}

\maketitle

\begin{abstract}
We study numerically the paramagnetic phase of the spin-1/2 
random transverse-field
Ising chain, using a mapping to non-interacting
fermions.  We extend our
earlier work, Phys. Rev.  {\bf 53}, 8486 (1996), to finite
temperatures and to dynamical properties. Our results are consistent with the
idea that there are
``Griffiths-McCoy'' singularities in the paramagnetic phase
described by a continuously varying
exponent $z(\delta)$, where $\delta$ measures the deviation from criticality.
There are some discrepancies between the values of $z(\delta)$ obtained from
different quantities, but this may be due to corrections to scaling. 
The {\em average} on-site
time dependent correlation function decays with a power law in the
paramagnetic phase, namely
$\tau^{-1/z(\delta)}$, where $\tau$ is imaginary time. However, the
{\em typical} value decays with a stretched exponential behavior,
$\exp(-c\tau^{1/\mu})$, where $\mu$ may be related to $z(\delta)$. 
We also obtain results for the full probability distribution of time
dependent correlation functions at different points in the paramagnetic phase. 

\end{abstract}

\vskip 0.3 truein
]

\section{Introduction}
Most critical points occur at a finite
temperature and one tunes through the transition by varying the temperature
itself. There are however, many critical points which occur at 
zero temperature and are traversed by varying
some other parameter. These quantum critical points
have recently been of great interest, particularly
for systems with disorder.
One reason for this interest is that even the
paramagnetic phase can have quite dramatic singularities. For example, in
systems with a discrete symmetry, such as the Ising model discussed
in this paper, there is a line in the phase diagram, on either side of
the critical point, where correlations in time (but not in space)
decay with a power law\cite{th,ry,gbh}.
Power law behavior is characteristic of a critical
point, but here this critical-like behavior occurs only in the time direction.
We could say that there is a line of ``semi-critical'' points. 
Furthermore, the exponent characterizing the power law decay is expected to
vary continuously along the line\cite{th,ry,gbh,dsf}.
As a result static response functions may
actually {\em diverge} in a finite region away from the critical
point\cite{mccoy,ry,gbh,dsf}.

These singularities arise from regions of the sample which have stronger
interactions than average and
were first discussed, many years ago, by Griffiths\cite{griffiths}, in the
context of {\em classical} models, where, however, they are rather
weak\cite{essen}. At
about the same time, McCoy\cite{mccoy}
determined exactly some properties of a two-dimensional classical
model (equivalent to the
disordered one-dimensional quantum magnet that we study here), finding that
the susceptibility diverges before the critical point is reached. It is now
understood that this behavior is due the
rare regions, more strongly coupled than average, discussed by Griffiths, but
which give a larger effect in the quantum regime than near a classical
transition. Hence we shall refer to Griffiths singularities in the
quantum regime as
Griffiths-McCoy singularities.

We shall study here the disordered spin-1/2 Ising chain in one-dimension, for
which many properties can be worked out in detail.
The ground state of this
model is closely related to the finite-temperature behavior of
a two-dimensional classical Ising model with disorder perfectly
correlated along one direction, which was first studied by McCoy~\cite{mccoy}
and by McCoy and Wu\cite{mw}.
Subsequently, the quantum model,
was studied by Shankar
and Murphy\cite{sm}, and, in great detail, by D.~S.~Fisher\cite{dsf}. From a
real space renormalization group analysis, which becomes exact on large
scales, Fisher obtained many new results and considerable physical
insight. One of the key conclusions is that many quantities have very broad
distributions, with average and typical values being quite different.
Consequently, a lot of information is lost by averaging.
Confirmation of Fisher's surprising predictions for the
$T=0$, equal time behavior
of the quantum problem, as well as some new
results for distributions of various quantities, were obtained in earlier
work\cite{yr} (henceforth denoted by YR),
which used a mapping of the spin problem to non-interacting
fermions\cite{lsm,pfeuty,katsura} to
obtain accurate numerical results for quite large systems.

In this paper we extend these techniques to finite temperatures and finite
times.  Our data suggest that the singularities are
governed by a continuously varying exponent, though there are some
discrepancies which we discuss.  We also obtain results
for the {\em distribution} of time dependent correlation functions, showing,
for example, that a {\em typical} correlation function decays with a
stretched exponential dependence on time, as opposed to the average which
decays with a power, as noted above.  Recently,
results for dynamical correlations at the critical point have been obtained by
Rieger and Igloi\cite{ri} using similar techniques to ours.  Hence our results
will be restricted to the paramagnetic phase.

\section{The Model}
The model that we study is 
one-dimensional random transverse-field Ising
chain with Hamiltonian
\begin{equation}
{\cal H} = -\sum_{i=1}^{L-1} J_i \sigma^z_i \sigma^z_{i+1} -
\sum_{i=1}^L h_i \sigma^x_i \ .
\label{ham}
\end{equation}
Here the $\{\sigma^\alpha_i\}$ are Pauli spin matrices, and the
interactions $J_i$ and transverse fields $h_i$ are both independent
random variables, with distributions $\pi(J)$ and $\rho(h)$ respectively.
The lattice size is $L$, and, in this paper, we will take {\em free},
rather than the more usual periodic, boundary
conditions. We will see later why this is necessary.

Since the model is in one-dimension, we
can perform a gauge transformation to make all the
$J_i$ and $h_i$ positive.
Unless otherwise stated,
the numerical work used the following rectangular distribution:
\begin{eqnarray}
\pi(J) & = & 
\left\{
\begin{array}{ll}
1 & \mbox{for $ 0 < J < 1$} \\
0  & \mbox{otherwise}
\end{array}
\right.
\nonumber \\ 
\rho(h) & = & 
\left\{
\begin{array}{ll}
h_0^{-1} & \mbox{for $ 0 < h < h_0$} \\
0  & \mbox{otherwise.}
\end{array}
\right.
\label{dist}
\end{eqnarray}
The model is therefore characterized by a
single control parameter, $h_0$. 
Defining
\begin{eqnarray}
\Delta_h & = & [\ln h]_{\rm av} \nonumber \\
\Delta_J & = & [\ln J]_{\rm av}
\end{eqnarray}
where $[\ldots]_{\rm av}$ denote an average over disorder, the
critical point occurs when\cite{sm,dsf}
\begin{equation}
\Delta_h = \Delta_J \ .
\end{equation}
Clearly this is satisfied if the distributions of bonds and fields are
equal, and the criticality of the model then follows from duality\cite{dsf}.
A convenient measure of the deviation from criticality is given by\cite{dsf}
\begin{equation}
\delta = { \Delta_h - \Delta_J \over 
\mbox{var}(\ln h) + \mbox{var} (\ln J) } \ ,
\label{delta}
\end{equation}
where $\mbox{var}(\cdots)$ denotes the variance. For the
distribution in Eq.~(\ref{dist}), we have
\begin{equation}
\delta = {1 \over 2} \ln h_0 \ .
\label{delta_h0}
\end{equation}

\section{The Method}
The numerical technique for static quantities, has been discussed in detail for
periodic boundary conditions
by YR. Here we consider free boundary conditions, which is simpler,
so will only summarize the main results. Following Lieb, Schultz and
Mattis\cite{lsm} we start by using the
the Jordan-Wigner transformation, which relates the
spin operators to fermion creation and annihilation operators,
$c^\dagger_i$ and $c_i$:
\begin{eqnarray}
\sigma^z_i & = & a^{\dagger}_i + a_i \nonumber \\
\sigma^y_i &  = & i( a^{\dagger}_i - a_i) \nonumber \\
\sigma^x_i &  = & 1 - 2 a^\dagger_i a_i =  1 - 2 c^\dagger_i c_i \ ,
\label{spins}
\end{eqnarray}
where
\begin{eqnarray}
a^\dagger_i & = & c^\dagger_i \exp\left[ -i\pi \sum_{j=1}^{i-1} c^\dagger_j c_j
\right] \nonumber \\
a_i & = & \exp\left[ -i\pi \sum_{j=1}^{i-1} c^\dagger_j c_j\right] c_i \ .
\label{jw}
\end{eqnarray}
The Hamiltonian, Eq.~(\ref{ham}), can then be written\cite{footnote2}
\begin{equation}
{\cal H} = \sum_{i=1}^L h_i (c_i^\dagger c_i - c_i c^\dagger_i)
- \sum_{i=1}^{L-1} J_i(c^\dagger_i - c_i)( c^\dagger_{i+1} + c_{i+1})  .
\label{hamfermi}
\end{equation}
Note that the fermion Hamiltonian, Eq~(\ref{hamfermi}), is
bi-linear and so describes {\em free} fermions.

We define operators $\Psi^\dagger_j$ for $1 \le j \le 2L$
by $\Psi^\dagger_i = c^\dagger_i$
and $\Psi^\dagger_{i+L} = c_i$, where
$1 \le i \le L$. Similarly $\Psi_i$ is the Hermitian conjugate of
$\Psi^\dagger_i$.
Note that the $\Psi_i$ and $\Psi^\dagger_j$ satisfy the usual fermion
commutation relations for {\em all} $i$ and $j$.

The Hamiltonian, Eq~(\ref{hamfermi}),
can then be written in terms of a real-symmetric
$2L \times 2L$ matrix, $\tilde{H}$, as
\begin{equation}
{\cal H} = \Psi^\dagger \tilde{H} \Psi \,
\end{equation}
where $\tilde{H}$ has the form 
\begin{equation}
\tilde{H} = \left[
\begin{array}{rr}
A & B \\
-B & -A
\end{array}
\right] \ ,
\label{htilde}
\end{equation}
where $A$ and $B$ are $L \times L$ matrices with elements 
\begin{eqnarray}
A_{i,i} & = & h_i \nonumber \\
A_{i,i+1} & =  - &J_i/2 \nonumber \\
A_{i+1,i} & =  - &J_i/2 \nonumber \\
B_{i,i+1} & =   &J_i/2 \nonumber \\
B_{i+1,i} & =  - &J_i/2 \ ,
\label{blocks}
\end{eqnarray}
where, since we have free boundary conditions, elements with an index $L+1$ are
zero. Note that $\tilde{H}$ is real symmetric.

Next we diagonalize $\tilde{H}$ numerically, 
to find the single particle eigenstates with eigenvalues
$\epsilon_\mu/2$, $\mu = 1, 2, \ldots 2 L$ and eigenvectors
$\Phi^\dagger_\mu$ which are linear combinations of the $\Psi^\dagger_i$
with real coefficients\cite{footnote1}.
It is easy to see that
the eigenstates come in pairs, with eigenvectors that are Hermitian
conjugates of each other and eigenvalues which are equal in magnitude
and opposite in sign. We can therefore define $\Phi^\dagger_\mu =
\gamma^\dagger_\mu$ if $\epsilon_\mu > 0$ and $\Phi^\dagger_{\mu^\prime} =
\gamma_\mu$ if $\mu^\prime$ is the state with energy $-\epsilon_\mu$.

The Hamiltonian can then be written just in terms of $L$ (rather than
$2L$) modes as
\begin{eqnarray}
{\cal H} & = &  \smfrac{1}{2}
\sum_{\mu=1}^L \epsilon_\mu (\gamma^\dagger_\mu \gamma_\mu -
\gamma_\mu \gamma^\dagger_\mu) \nonumber \\
& = & \sum_{\mu=1}^L \epsilon_\mu
(\gamma^\dagger_\mu \gamma_\mu - \smfrac{1}{2}) ,
\label{hamdiag}
\end{eqnarray}
where all the $\epsilon_\mu$ are now taken to be positive. The average energy
per site is therefore given by
\begin{equation}
E_{\rm av}  = {1\over L} \sum_{\mu=1}^L
[ \epsilon_\mu (n_\mu -\smfrac{1}{2})]_{\rm av},
\end{equation}
\label{energy}
where $n_\mu$ is the Fermi function,
\begin{equation}
n_\mu = {1 \over \exp(\beta \epsilon_\mu) + 1 } ,
\end{equation}
and the specific heat is consequently given by
\begin{equation}
C_{\rm av} = {1\over L T^2} \sum_{\mu=1}^L
[ \epsilon_\mu^2 
n_\mu (1 - n_\mu) ]_{\rm av},
\label{spec_heat}
\end{equation}
in units where $k_B=1$, which we take from now on.

We next consider equal time correlation functions, defined by
\begin{equation}
S_{ij} = \langle \sigma^z_i \sigma^z_j \rangle \ .
\end{equation}
Since $S_{ji}=S_{ij}$, we can take $j > i$, without loss of generality. 
$S_{ij}$ is given in terms of a determinant of size $j-i$ by\cite{lsm}
\begin{equation}
S_{ij} =
\left|
\begin{array}{cccc}
G_{i, i+1} & G_{i, i+2} & \cdots & G_{ij} \\
G_{i+1, i+1} & G_{i+1, i+2}  & \cdots &  G_{i+1,j} \\
\vdots & \vdots & \ddots & \vdots \\
G_{j-1, i+1} & G_{j-1, i+2}& \cdots & G_{j-1, j}
\end{array}
\right| \ ,
\label{det}
\end{equation}
where
\begin{equation}
G_{ij} = \langle (c^\dagger_i - c_i) (c^\dagger_j + c_j) \rangle.
\end{equation}

$G_{ij}$ can be expressed in
terms of the eigenvectors of the matrix $\tilde{H}$
in Eq.~(\ref{htilde}). Let us write 
\begin{eqnarray}
c^\dagger_i + c_i & = & \sum_{\mu=1}^L
\phi_{\mu i} (\gamma^\dagger_\mu + \gamma_\mu)
\nonumber \\
c^\dagger_i - c_i & = & \sum_{\mu=1}^L
\psi_{\mu i} (\gamma^\dagger_\mu - \gamma_\mu)
\ ,
\end{eqnarray}
where $\psi$ and $\phi$ can be shown to be orthogonal matrices. Then
\begin{eqnarray}
G_{ij} & = & \langle (c^\dagger_i - c_i) (c^\dagger_j + c_j) \rangle
\nonumber \\
 & = & \sum_{\mu=1}^L \psi_{\mu i} \phi_{\mu j}
 \langle (\gamma^\dagger_\mu - \gamma_\mu) (\gamma^\dagger_\mu + \gamma_\mu)
 \rangle
\nonumber \\
 & = &  -\sum_{\mu=1}^L (\psi^T)_{i\mu}(1 - 2n_\mu)
 \phi_{\mu j} \ ,
\label{G}
\end{eqnarray}
since
\begin{equation}
\langle \gamma^\dagger_\mu \gamma^\dagger_\mu \rangle =
\langle \gamma_\mu \gamma_\mu \rangle = 0 
\end{equation}
\begin{equation}
\langle \gamma^\dagger_\mu \gamma_\mu \rangle = 1 - \langle \gamma_\mu
\gamma^\dagger_\mu \rangle = n_\mu .
\end{equation}
At zero temperature, Eq.~(\ref{G}) goes over to Eq.~(54) of YR.

We now discuss how these results are generalized to time dependent correlation
functions\cite{sy,snm}. We are interested in the $\sigma^z$--$\sigma^z$
imaginary time correlation function, $S_{ij}(\tau)$ for $0 \le \tau \le \beta$,
where
\begin{equation}
S_{ij}(\tau_1 - \tau_2) = \langle \sigma^z_i(\tau_1) \sigma^z_j(\tau_2)
\rangle,
\end{equation}
with
\begin{equation}
\sigma^z_i(\tau_1) = e^{\tau_1{\cal H}} \sigma^z_i e^{-\tau_1{\cal H}}.
\end{equation}
Note that 
\begin{equation}
S_{ij}(\tau) = S_{ji}(\beta - \tau) ,
\label{ijji}
\end{equation}
which follows by cyclically permuting the trace, and 
so, without loss of generality, we just consider $j \ge i$.

Substituting the transformation in Eq.~(\ref{jw}) one has
$$
S_{ij}(\tau) = \left\langle 
\exp\left[ -i\pi \sum_{m=1}^{i-1} c^\dagger_m(\tau) c_m(\tau) \right]
\left(c^\dagger_i(\tau) + c_i(\tau) \right) \right.
$$
\begin{equation}
\left.
\times \exp\left[ -i\pi \sum_{l=1}^{j-1} c^\dagger_l c_l \right]
\left(c^\dagger_j + c_j \right) \right\rangle .
\end{equation}
This can be simplified since
\begin{equation}
\exp \left[-i\pi c^\dagger_m c_m \right] = A_m B_m
\label{AB}
\end{equation}
where
\begin{eqnarray}
A_m & = & c^\dagger_m + c_m \nonumber \\
B_m & = & c^\dagger_m - c_m ,
\end{eqnarray}
and so
\begin{equation}
S_{ij}(\tau) = \left\langle 
[ \prod_{m=1}^{i-1} A_m(\tau)B_m(\tau) ] A_i(\tau) 
[ \prod_{l=1}^{j-1} A_l B_l ] A_j  \right\rangle .
\label{ctau}
\end{equation}
Hence $S_{ij}(\tau)$ involves the product of $2(i+j-1)$ Fermi operators. 
This number is very large if one
is interested in two sites in near the center of a large lattice, even if those
sites are close together.

The situation is much simpler for $t=0$ because then the product of
all the operators to the left of site $i$ is unity\cite{footnote4}, and so
\begin{eqnarray}
S_{ij}(0) & = & \left\langle 
A_i [ \prod_{l=i}^{j-1} A_l B_l ] A_j  \right\rangle \nonumber \\
& = &\left\langle B_i [ \prod_{l=i+1}^{j-1} A_l B_l ] A_j  \right\rangle ,
\end{eqnarray}
where the last line follows because $A_i^2=1$. As shown by Lieb et al.\cite{lsm},
Wick's theorem, together with the observation that
\begin{equation}
\langle A_i A_j \rangle = -\langle B_i B_j\rangle = \delta_{ij} ,
\label{aabb}
\end{equation}
enables one to write $S_{ij}(t=0)$ as the Toeplitz determinant of order $j-i$
in Eq.~(\ref{det}). This is convenient because the determinant
is small if $i$ and $j$
are close together, even for an sites far from the boundary of a large system.

For $\tau \ne 0$ in Eq.~(\ref{ctau}) one can still use Wick's theorem but
now there are many more pairs of operators to be included in the contractions.
Wick's theorem for fermions requires the sum over all possible products of
pair-averages, with a sign which is 1 or $-1$ depending on whether an even or
odd permutation of the operators is necessary to get the operators in the
product back to the original order. This is called a Pfaffian, see e.g. the
book by McCoy and Wu\cite{mwbook}. If the number of operators, $2n$, 
is large, (here $n=i+j-1$), evaluation of the Pfaffian is intractable
because the number of terms, $(2n-1)!!$, is too large even for fast computers.
However, the Pfaffian is also
the square root of an antisymmetric matrix (of order $2n$) formed from the
pair averages. To be precise, if $A, B, \cdots , Z$ are Fermi operators, and
the average is over a free Fermi Hamiltonian (in the grand canonical ensemble),
one has\cite{mwbook}
\begin{equation}
\langle ABC\cdots Z\rangle = 
\left|
\begin{array}{ccccc}
0 & \langle AB \rangle & \langle AC \rangle & \cdots &\langle AZ \rangle  \\
-\langle AB \rangle & 0 & \langle BC \rangle & \cdots &\langle BZ \rangle \\
-\langle AC \rangle & -\langle BC \rangle & 0 & \cdots &\langle CZ \rangle \\
\vdots & \vdots & \vdots & \ddots & \vdots \\
-\langle AZ \rangle & -\langle BZ \rangle & -\langle CZ \rangle & \cdots &0
\end{array}
\right|^{1/2}.
\label{abcz}
\end{equation}

As a simple example, consider $i=1, j=2$ in Eq.~(\ref{ctau}) for which Wick's
theorem gives
\begin{eqnarray}
S_{1,2}(\tau) & = & \langle A_1(\tau) A_1 B_1 A_2 \rangle \nonumber \\
& = & \langle A_1(\tau) A_1 \rangle  \langle B_1 A_2 \rangle -
\langle A_1(\tau) B_1 \rangle  \langle A_1 A_2 \rangle  \nonumber \\
& & \qquad + \langle A_1(\tau) A_2 \rangle  \langle A_1 B_1 \rangle  .
\label{example_wick}
\end{eqnarray}
This is easily shown to equal 
\begin{equation}
\left|
\begin{array}{cccc}
0 & \langle A_1(\tau)A_1 \rangle & \langle A_1(\tau)B_1 \rangle &
 \langle A_1(\tau)A_2 \rangle \\
-\langle A_1(\tau)A_1 \rangle & 0 & \langle A_1 B_1 \rangle &
 \langle A_1 A_2 \rangle \\
-\langle A_1(\tau)B_1 \rangle & -\langle A_1 B_1 \rangle & 0 &
 \langle B_1 A_2 \rangle \\
-\langle A_1(\tau)A_2 \rangle & -\langle A_1 A_2 \rangle & -\langle B_1 A_2
\rangle & 0 \\
\end{array}
\right|^{1/2},
\end{equation}
which is just Eq.~(\ref{abcz}) for this case.
Note that Eq.~(\ref{aabb}) gives $\langle A_1 A_2\rangle = 0$ and we also have
$A_1^2=1$.
Hence, for
$t=0$, we get, from Eq.~(\ref{example_wick}), the simpler result 
\begin{equation}
S_{1,2}(0) = \langle B_1 A_2 \rangle ,
\end{equation}
which also follows immediately from Eq.~(\ref{det}).

It is much more convenient to work evaluate the determinant and take the square
root, than to directly evaluate the Pfaffian. In numerical work, the number
of operations required to evaluate the determinant is of order $(2n)^3$, which,
for $n$ a few hundred,
is feasible, whereas the $(2n-1)!!$ operations to
evaluate the Pfaffian is definitely not.

The pair averages needed for Wick's theorem can be evaluated in the
same manner
used to derive Eq.~(\ref{G}) above for the equal time correlation functions.
The result is,
\begin{eqnarray}
\langle A_i(\tau_1) B_j(\tau_2) \rangle & = &
\sum_{\mu=1}^L \left(\phi^T\right)_{i\mu}
\left[
-U_\mu(\tau) + V_\mu(\tau) 
\right] \psi_{\mu j} \nonumber \\
\langle B_i(\tau_1) A_j(\tau_2) \rangle & = &
\sum_{\mu=1}^L \left(\psi^T\right)_{i\mu}
\left[
U_\mu(\tau) - V_\mu(\tau) 
\right] \phi_{\mu j} \nonumber \\
\langle A_i(\tau_1) A_j(\tau_2) \rangle & = &
\sum_{\mu=1}^L \left(\phi^T\right)_{i\mu}
\left[
U_\mu(\tau) + V_\mu(\tau) 
\right]
\phi_{\mu j}  \nonumber\\
\langle B_i(\tau_1) B_j(\tau_2) \rangle & = &
\sum_{\mu=1}^L \left(\psi^T\right)_{i\mu}
\left[
-U_\mu(\tau) - V_\mu(\tau) 
\right]
\psi_{\mu j} , \nonumber \\
\end{eqnarray}
where
\begin{eqnarray}
U_\mu(\tau)& = & n_\mu e^{\epsilon_\mu\tau} \nonumber \\
V_\mu(\tau)& = & (1-n_\mu)e^{-\epsilon_\mu\tau} ,
\end{eqnarray}
and $\tau=\tau_1 - \tau_2$. 
From these pair averages the
determinant in Eq.~(\ref{abcz}) [with the operators in Eq.~(\ref{ctau})]
is evaluated numerically, and finally the square root
taken. Since the imaginary time correlation function is real and
positive, there is no ambiguity about the sign.

We concentrate on the {\em local} correlation function,
$S_{ii}(\tau)$.
This determines 
the local susceptibility from
\begin{equation}
\chi_{ii} = \int_0^\beta S_{ii}(\tau) \, d\tau .
\label{integral}
\end{equation}
We determine $S_{ii}(\tau)$ for different values of $\tau$,
increasing in a roughly logarithmic manner, and approximate the integral by
the trapezium rule.
We compute the average local correlation function and average local
susceptibility, defined by
\begin{equation}
S_{\rm av}(\tau) = {1\over L} \sum_i [ S_{ii}(\tau) ]_{\rm av} ,
\qquad 
\chi^{\rm loc}_{\rm av} = {1\over L} \sum_i [ \chi_{ii}]_{\rm av} .
\end{equation}
From now on, for compactness of notation, we will denote $S_{ii}(\tau)$ by
$S(\tau)$.
In addition, because
there are large fluctuations in the values of
$S(\tau)$
from site to site, 
we also look at the distributions of this quantity for different $\tau$.

\section{Phenomenological Description}
\label{phenon}
From recent work,\cite{th,ry,gbh,dsf,yr,ri}
a phenomenological description of the Griffiths-McCoy region of the
paramagnetic
phase has emerged. Singularities arise from small regions which are ``locally
in the ferromagnetic phase'' and have a very small energy gap. As a result,
there are low energy ``cluster'' excitations, which have a power law
distribution of energies, $\Delta E$.
The probability of having a low energy excitation is
proportional to the size, $L$, and so we can write the dimensionless
probability,
$\Delta E\, P(\Delta E)$,as
\begin{equation}
\Delta E\, P(\Delta E) \sim L (\Delta E)^{1/z(\delta)} ,
\label{dist_de}
\end{equation}
where we write the power in terms of
a dynamical exponent $z(\delta)$, since Eq.~(\ref{dist_de}) corresponds to the
standard relation between a length scale ($L$ here) and an energy scale
($\Delta E$) here. The exponent is expected to vary continuously in the
Griffiths-McCoy phase, and we indicate this by the notation, $z(\delta)$.

For the distribution used here, Eq.~(\ref{dist}),
in the limit $\delta \to \infty$
we can neglect the interactions and the excitation energy of a single spin
is just $2 h_i$. Since the $h_i$
have a uniform distribution for $h \to 0$, it follows that $z=1$ in this limit,
i.e.
\begin{equation}
\lim_{\delta\to\infty} z(\delta) = 1.
\label{delta_inf}
\end{equation}
Furthermore, it has been established\cite{dsf,sm}
that $z(\delta)\to\infty$  at the critical
point, i.e.
\begin{equation}
\lim_{\delta\to 0} z(\delta) = \infty.
\end{equation}

The energy per site
at low temperature can be
estimated from the energy of the excited clusters, i.e.
\begin{eqnarray}
E_{\rm av}(T) - E_{\rm av}(0) & \approx & {1\over L}\int \Delta E\, P(\Delta E) 
{\exp{(-\beta\Delta E)} \over 1 + \exp{(-\beta\Delta E)} } \nonumber \\
& \sim & T^{1 + 1/z(\delta)} ,
\end{eqnarray}
so the specific heat varies as
\begin{equation}
C_{\rm av} \sim  T^{1/z(\delta)} ,
\label{shav}
\end{equation}
for $T \to 0$.

For a site in a cluster with a low energy excitation $\Delta E$, the long time
behavior of $S(\tau)$ is $\exp(-\Delta E \tau)$. Averaging over the
distribution gives
\begin{equation}
S_{\rm av}(\tau) \sim {1 \over \tau^{1/z(\delta)} } ,
\label{ctauav}
\end{equation}
and integrating $\tau$ up to $\beta$ gives
\begin{equation}
\chi^{\rm loc}_{\rm av} \sim T^{ -1 + 1/ z(\delta)} ,
\label{chiav}
\end{equation}
for $T \to 0$.

To our knowledge, not much is known about the distribution of $S(\tau)$
and one of the goals of the present work is to deduce its form. Note that
because we expect the distribution to be very broad, with average and typical
values quite different, we cannot predict its form simply from knowing
the average.

\section{Results}
\begin{figure}
\epsfxsize=\columnwidth\epsfbox{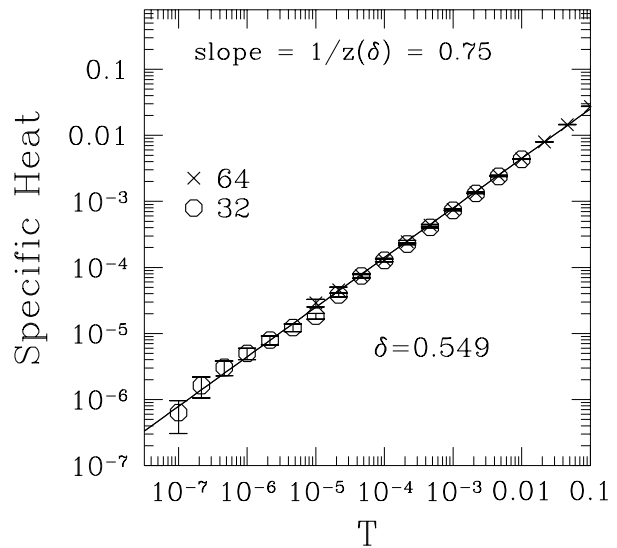}
\caption{
The specific heat as a function of $T$ for $\delta = 0.549\ (h_0 = 3)$
and sizes $L=64$ and 32.
According to
Eq.~(\protect\ref{shav}),
the slope is $1/z(\delta)$.
}
\label{sh_3}
\end{figure}
\begin{figure}
\epsfxsize=\columnwidth\epsfbox{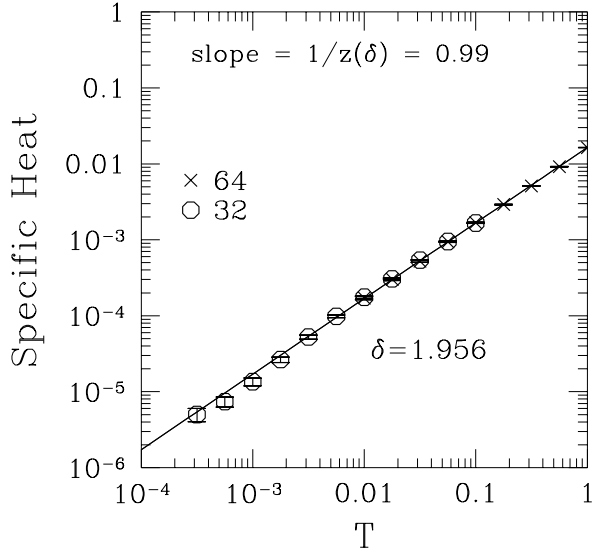}
\caption{
The specific heat as a function of $T$ for $\delta = 1.956\ (h_0 = 50)$
and sizes $L=64$ and 32.
According to
Eqs.~(\protect\ref{shav}) and (\protect\ref{delta_inf}),
the slope should be close to unity, as
indeed it is.
}
\label{sh_50}
\end{figure}

We start with our results for the specific heat. The data for $h_0=3
\ (\delta=0.549)$, shown in
Fig.~\ref{sh_3}, has
good straight line behavior for more than five decades,
the slope, equal to $1/z(\delta)$, is 0.74.

According to Eq.~(\ref{delta_inf}) $z(\delta)$ should tend to unity as $\delta
\to\infty$. This is confirmed by the data for the specific
heat for $h=50\ (\delta= 1.956)$,
shown in Fig.~\ref{sh_50}, which has a slope of 0.99, very
close to the expected value.

\begin{figure}
\epsfxsize=\columnwidth\epsfbox{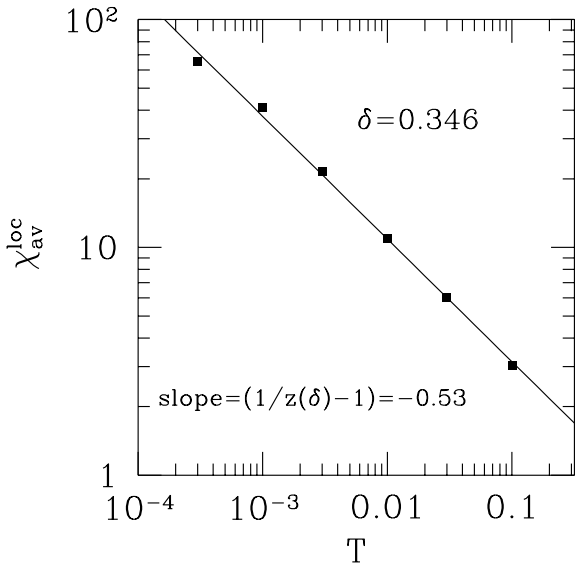}
\caption{
The susceptibility heat as a function of $T$ for $\delta = 0.346$
and size $L=64$.
}
\label{chi_2}
\end{figure}

The temperature dependence of the susceptibility is shown in Fig.~\ref{chi_2}
for $h_0= 2 \ (\delta=0.346)$.
Again a power law behavior is obtained, as expected from
Eq.~(\ref{chiav}).

\begin{figure}
\epsfxsize=\columnwidth\epsfbox{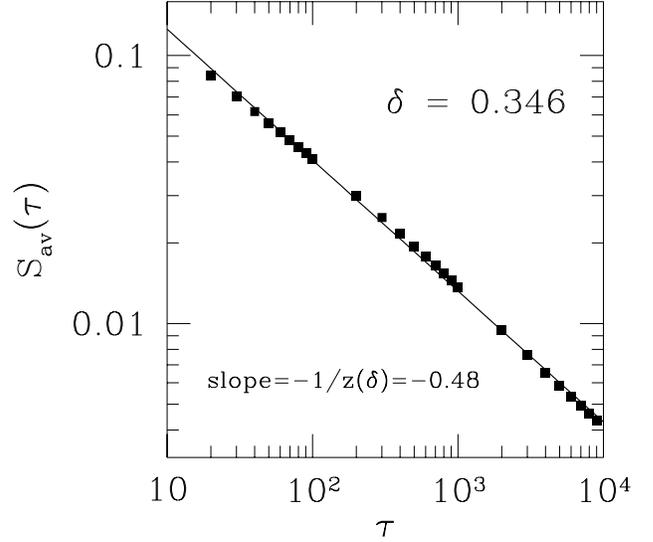}
\caption{
The average on-site (imaginary) time dependent correlation function at T=0 for
$\delta=0.346$. The lattice size is $L=64$.
}
\label{ctau_2}
\end{figure}

\begin{figure}
\epsfxsize=\columnwidth\epsfbox{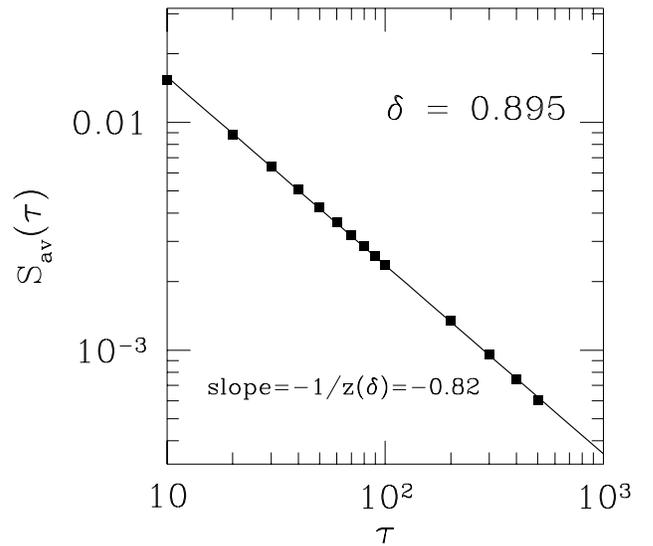}
\caption{
As for Fig.~\protect\ref{ctau_2}, but with $\delta=0.895$.
}
\label{ctau_6}
\end{figure}

We now proceed to our results for the time dependent correlation functions. 
Data for $S_{\rm av}(\tau)$ for $h_0=2$ and 6 ($\delta=0.346$ and 0.895)
at $T=0$
are shown in Figs.~\ref{ctau_2} and \ref{ctau_6} on a
double logarithmic scale. The results clearly indicate a power law behavior, as
expected from Eq.~(\ref{ctauav}).


\begin{figure}
\epsfxsize=\columnwidth\epsfbox{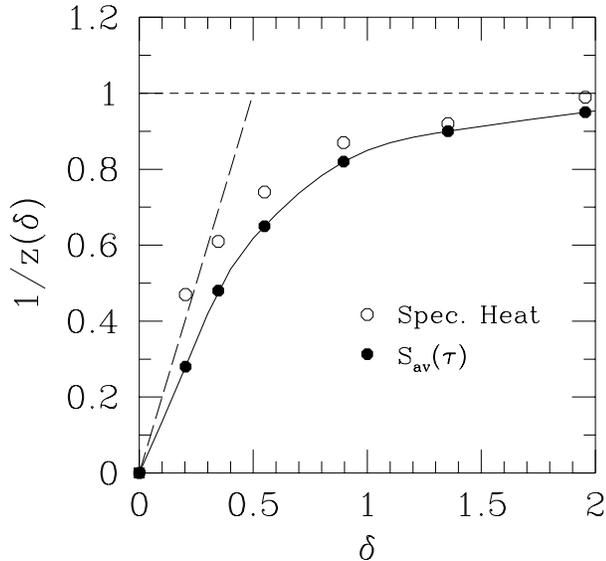}
\caption{
The dynamic exponent $z(\delta)$ obtained both from the specific heat and the
time dependent correlation function.
The short dashed line indicates the
asymptotic value $z(\delta\to\infty)=1$, and the long dashed line is the
prediction of Fisher\protect\cite{dsf}, $z^{-1}=2\delta$ for $\delta\to 0$. At
the critical point, $\delta=0$, $z(\delta)$ is predicted to be infinite. 
}
\label{z}
\end{figure}
From our results for $C_{\rm av}$ and $S_{\rm av}(\tau)$, we obtain
$z(\delta)$ from Eqs.~(\ref{shav}) and (\ref{ctauav}). We
summarize\cite{footnote5}
these results in Fig.~\ref{z}.
While the trend in the two sets of data is the same, there are some
differences, which we do not understand very well. 
The lattice sizes used are quite large and the data used to generate the
estimates for $z(\delta)$ fit
a straight line over a fairly large range, especially for the
specific heat. 
This suggests that the estimates for $z(\delta)$ should agree well.
However, corrections to scaling
may be large because free boundary conditions are used here, as opposed to
the more conventional periodic boundary conditions. 

\begin{figure}
\epsfxsize=\columnwidth\epsfbox{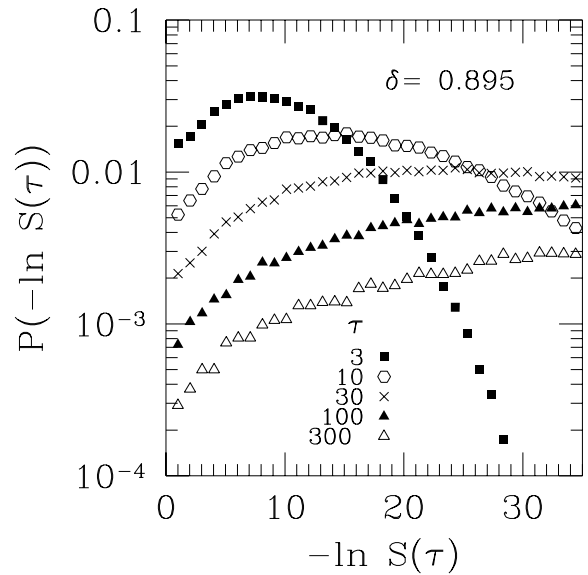}
\caption{
The distribution of $-\ln S(\tau)$ for $\delta=0.895$.
The lattice size is $L=64$ and
$50,000$ samples were averaged over. 
}
\label{plnc_6}
\end{figure}
So far we have just considered the {\em average} value of various quantities.
However, one of the most surprising features of this model is that 
distributions are so broad that average and typical values can be quite
different. We have therefore also studied the {\em distribution} of the on-site
time dependent correlation functions for different values of $\tau$.

\begin{figure}
\epsfxsize=\columnwidth\epsfbox{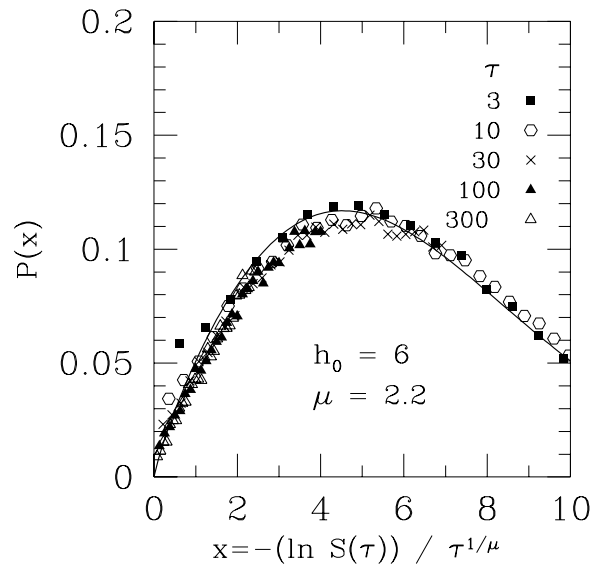}
\caption{
A scaling plot of the data in Fig.~\protect\ref{plnc_6}. The scaling variable
is given in Eq.~(\protect\ref{scale_x})
and we have taken $\mu=2.2$ to get the best
fit. The solid line is Eq.~(\protect\ref{scale_dist}) with $z=1.2$,
consistent with Eq.~(\protect\ref{mu}), and $c=0.2$. The fit to the data is
good.
}
\label{plnc_scale_6}
\end{figure}

Fig.~\ref{plnc_6} shows the distribution of $\ln S(\tau)$ for $\delta=0.895$.
The
distribution is broad and becomes broader with increasing $\tau$. We therefore
attempt to collapse the data by plotting it versus the scaling variable
\begin{equation}
x = -{\ln S(\tau) \over \tau^{1/\mu} }
\label{scale_x}
\end{equation}
with an appropriate choice for $\mu$.
We shall see that, at a given value of $\delta$, the distributions of $S(\tau)$
for different times are all given by a {\em single} scaling function $P(x)$.

\begin{figure}
\epsfxsize=\columnwidth\epsfbox{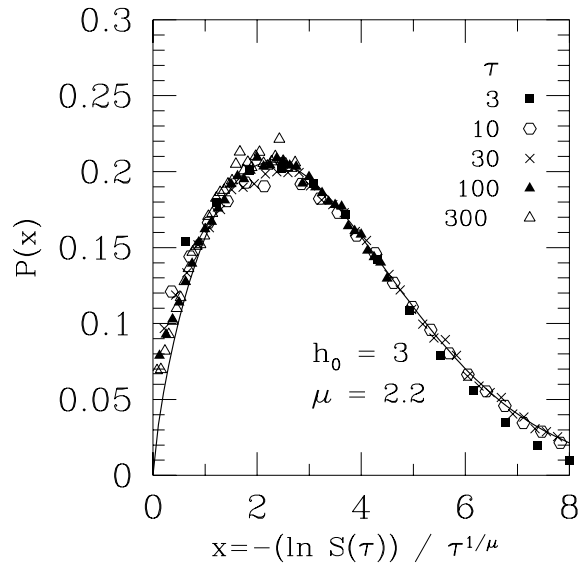}
\caption{
As for Fig.~\protect\ref{plnc_scale_6} but with $\delta = 0.549$.
The best fit is for $\mu=2.2$. The solid line corresponds to the scaling
distribution in Eq.~(\protect\ref{scale_dist}) with $z(\delta)=1.36$
and the scale
factor $c=0.36$. Although the fit is not too bad, it is actually somewhat
inconsistent because $z(\delta)$ and $\mu$ should be related by
Eq.~(\protect\ref{mu}), which would require $z=1.2$. However, this choice of
$z(\delta)$ works less well.
}
\label{plnc_scale_3}
\end{figure}

\begin{figure}
\epsfxsize=\columnwidth\epsfbox{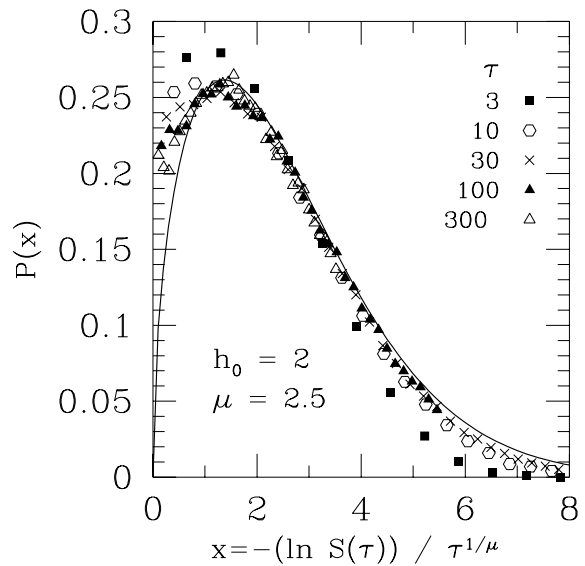}
\caption{
As for Fig.~\protect\ref{plnc_scale_6} but with $\delta=0.346$. The best fit is
for $\mu=2.5$.  The solid line corresponds to the scaling
distribution in Eq.~(\protect\ref{scale_dist}) with $z(\delta)=2.0$
and the scale factor $c=0.46$. However, this is inconsistent since,
according to the theory in the
appendix, $\mu$ and $z(\delta)$ should be related by 
Eq.~(\protect\ref{mu}), which would require $z=1.5$. However, this choice of
$z(\delta)$ works less well.
}
\label{plnc_scale_2}
\end{figure}

Fig.~\ref{plnc_scale_6}
shows that the data plotted in
this way collapses quite well with the choice
$\mu=2.2$.
For small $x$ the behavior seems to be close to linear, though
the data for small
times has a finite intercept, which perhaps vanishes for $\tau\to\infty$. We
shall discuss this again later. Scaling plots for
$\delta=0.549$ and 0.346 are shown in
Figs.~\ref{plnc_scale_3} and \ref{plnc_scale_2}.
The values of $\mu$ used in these fits are 2.2 and 2.5.

From a variable range hopping analysis, 
M.~P.~A.~Fisher\cite{mpaf:pc} has argued that $\mu$ should be related to
$z(\delta)$ by
\begin{equation}
\mu = 1 + z(\delta) .
\label{mu}
\end{equation}
In Appendix A we show that, within the same set of assumptions,
the scaling function is given by
\begin{equation}
P(x) = c (cx)^{1/z} \exp\left[-\left({z \over z +1}\right) (cx)^{(z+1)/z} 
\right]  ,
\label{scale_dist}
\end{equation}
where
\begin{equation}
x = {-\ln S(\tau) \over \tau^{1/(z+1)} } ,
\label{x_scale}
\end{equation}
as also follows from Eqs.~(\ref{scale_x}) and (\ref{mu}),
and $c$ is a scale factor.
The data for $\delta=0.895$ fits the scaling function in
Eq.~(\ref{scale_dist}) quite well, with an appropriate choice of the scale
factor $c$ and a value for $z(\delta)$, and hence $\mu$, consistent with that
from other data shown in Fig.~\ref{z}.
However, the data closer to the critical point in
Figs.~\ref{plnc_scale_3} and \ref{plnc_scale_2} only fit
the scaling form
if $z(\delta)$ in Eq.~(\ref{scale_dist}) and $\mu$ in Eq.~(\ref{scale_x})
are adjusted independently.
The values for $z(\delta)$
are found to be in reasonable agreement with those in Fig.~\ref{z}, but
the values for $\mu$ are then somewhat inconsistent with Eq.~(\ref{mu}).
Note however, that the 
stretched exponential is only expected\cite{mpaf:pc} to be valid
for times larger than a characteristic
time which diverges at the transition. Hence it is possible that our data is
not at sufficiently long times for smaller $\delta$ to get a good estimate for
$\mu$..


The fact that the scaling variable is given by Eq.~(\ref{scale_x}) shows that
the {\em typical} correlation function, which we define to be the exponential of
the average of the log, is given by
\begin{equation}
S_{\rm typ}(\tau) \equiv \exp([\ln S(\tau)]_{\rm av}) \sim
\exp(-c\tau^{1/\mu}), 
\label{stretched}
\end{equation}
where $c$ is a constant,
i.e. a stretched exponential. This contrasts with the average, which varies as
a power of $\tau$, as shown in Eq.~(\ref{ctauav}).

The average value is dominated by rare regions which have a much larger
correlation function at long times than in a typical region. It is
interesting to ask whether the average value is contained within 
the scaling function,
or whether it comes from contributions
which are actually corrections to scaling. We
shall argue that there are 
{\em both} scaling and and non-scaling
contributions to the $\tau^{1/z(\delta)}$ behavior in Eq.~(\ref{ctauav}).
Similar behavior
has been found recently for the distribution of the equal-time end-to-end
correlation function at the critical point\cite{fy}.

To estimate the contribution from the scaling function, we assume that the
distribution, $P(x)$, of the scaling variable $x$, in Eq.~(\ref{scale_x}), has
the form $x^a$ in the limit $x\to 0$, the only region
which contributes to the
average. This is indeed the case for the phenomenological theory discussed in
the appendix, for which $P(x)$ is given in
Eq.~(\ref{scale_dist}).
Since $S(\tau) = \exp(-x\tau^{1/\mu})$, it follows that the scaling
contribution to the average is
\begin{eqnarray}
S_{\rm av}(\tau) &\sim & \int_0^\infty  x^a \exp(-x\tau^{1/\mu})
\, dx\nonumber\\
& \sim & {1 \over \tau^{(1+a) / \mu} } ,
\end{eqnarray}
which is of the form in Eq.~(\ref{ctauav}) with the identification
\begin{equation}
z(\delta) = {\mu \over 1 + a} .
\label{zmu}
\end{equation}
Note that the phenomenological theory,
Eqs.~(\ref{mu})--(\ref{x_scale}), satisfies this condition with
\begin{equation}
a = {1 \over z(\delta) }.
\end{equation}
Although the numerical data 
only fits the phenomenological
theory close to the critical point with values of $\mu$ and $z(\delta)$ which
are inconsistent with Eq.~(\ref{mu}), this may be due to the times studied not
being long enough and other corrections to scaling. It seems very likely that
there {\em is} a contribution to the average time dependent correlation
function from the scaling function. 

\begin{figure}
\epsfxsize=\columnwidth\epsfbox{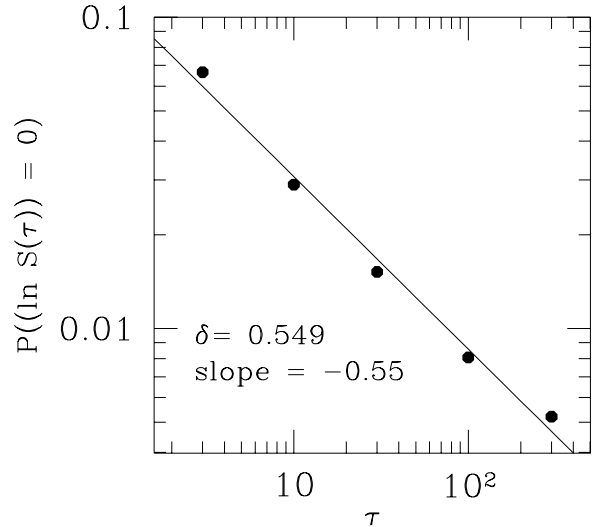}
\caption{
The probability of $y= -\ln S(\tau)$ at $y=0$ for different values of $\tau$
for $\delta=0.549$ and $L=64$.
The slope is $-0.55$ which is somewhat less than, but not
too far from the
value of $1/z(\delta)$ in Fig.~\protect\ref{z}. Hence this finite value
of the probability at $\ln S(\tau)=0$ (which is a correction to the scaling
form) gives a contribution to the power law decay of the average correlation
function shown in Eq.~(\protect\ref{ctauav}).
}
\label{ps0eq0}
\end{figure}

However, we shall now argue that there is an additional contribution to the
average correlation function, which comes from {\em corrections} to scaling.
The exponent $a$ is positive, so
$P(0)=0$
in the scaling limit. However, we see from the data in Fig.~\ref{plnc_6} that
the value of ${\rm Prob}\ (-\ln S(\tau) = 0)$
is finite for finite $\tau$ though it does
decrease for large $\tau$. This finite intercept is a correction to scaling.
It is easy to see that if ${\rm Prob}\ (-\ln S(\tau) = 0)
\sim \tau^{-1/z(\delta)}$,
then this finite intercept will give a power law 
contribution to the average correlation function
of the form in
Eq.~(\ref{ctauav}). Results for ${\rm Prob}\ (-\ln S(\tau) = 0)$
for different values of
$\tau$ at $\delta=0.549$
are shown in Fig.~\ref{ps0eq0}. The slope is $-0.55$, which
is a little less than the estimate for $1/z(\delta)$ 
obtained by other methods, see Fig.~\ref{z}, but
since
there are errors in extrapolating the data to $S(\tau)=0$,
this is perhaps not significant. 
Hence our results suggest that
${\rm Prob}\ (-\ln S(\tau) = 0)$
vanishes as $\tau^{-1/z(\delta)}$, which
gives an additional
contribution to the average correlation function which is not contained
within the scaling function. 

\section{Conclusions}
We have studied the paramagnetic phase of the
disordered one-dimensional Ising chain in a
transverse field. 
{\em Average} properties (equal time, time dependent,
and temperature dependent) 
can be characterized by a continuously varying exponent
$z(\delta)$. At criticality, $z(0)=\infty$ and, for the model studied,
$\lim_{\delta\to\infty} z(\delta)=1$.
As an example, the average, on-site, time-dependent correlation function decays
with a continuously varying exponent, see Eq.~(\ref{ctauav}). There are some
discrepancies between the values of $z(\delta)$ obtained in different ways. We
suspect that they are due to corrections to scaling, but one should perhaps
also worry that the simple phenomenological picture in Sec.~\ref{phenon} may be
inadequate.

By contrast, the
{\em typical} time dependent
correlation function decays with the stretched exponential form in
Eq.~(\ref{stretched}),
where the exponent $\mu$ is probably given by Eq.~(\ref{mu}), though our data
close to the critical point shows some discrepancy with this.
At long times,
the distribution of $S(\tau)$ has a scaling form, being
a function, $P(x)$, of the single variable
$x$ in Eq.~(\ref{scale_x}). This may be given by Eq.~(\ref{scale_dist}), though
corrections to scaling in the small $x$ region prevent us from checking this
thoroughly. For example,
the data in the scaling
plots in Figs.~\ref{plnc_scale_6}-\ref{plnc_scale_2}
are expected to go through the origin, but they
actually do not, the discrepancy being larger for smaller $\tau$.

The average of $S(\tau)$ arises both from the small
$x$ region of the scaling function, and from corrections to scaling.

It is interesting to ask to what extent the behavior in the paramegnetic phase
is universal. Presumably the dependence of $z$ on $\delta$ is non-universal
(except close to the critical point\cite{dsf}) but, for a given z,
is the scaling function for the ditribution of local correlation
functions universal? Since the scaling function involves the limit of long
times, it is possible that the microscopic details do not matter, only the
form of the low energy density of states, in which case the distribution would
be universal.

It would be interesting to see to what extent the results found here
in one dimension go over to higher dimensions.

\acknowledgments
I should like to thank D.~S.~Fisher and M.~P.~A.~Fisher
for many stimulating comments. I should also like to thank G.~Kotliar for a
helpful discussion.
This work is supported by the National Science
Foundation under grant No. DMR--9411964. 

\appendix
\section{}
Following the suggestion of
M.~P.~A.~Fisher,\cite{mpaf:pc}
we assume a simple phenomenological picture in which there are
cluster excitations localized about different sites .
It is plausible that the wavefuntion of an excitation will
decay exponentially with a distance of order the typical correlation function,
$\xit$.
The contribution to the dynamics of site at the origin from
an excitation centered
$n$ lattice spacings away will therefore be
$\exp\left(-r / \xit - \epsilon_n \tau\right)$.
The local correlation function is therefore given by
\begin{equation}
S(\tau) =  A \sum_{n=0}^\infty
\exp\left(-{r\over \xit} - \epsilon_n \tau\right) ,
\label{sum}
\end{equation}
where 
\begin{equation}
A = 1 - e^{1/\xi} \qquad ( \simeq \xi^{-1} \ \mbox{for} \ \xi \gg 1 ) ,
\end{equation}
ensures that $S(0) = 1$.

As a simple model we will assume that the
$\epsilon_n$ are uncorrelated. 
Writing $\epsilon_n$ = $\epsilon_0 e_n$, where $\epsilon_0$ is a
characteristic energy scale, we take the distribution of the $e_n$ to be
\begin{equation}
\rho(e) = (\lambda+1)e^{\lambda} , \qquad (0 < e < 1 ) ,
\end{equation}
where we see from Eq,~(\ref{dist_de}) that
\begin{equation}
\lambda = {1\over z(\delta)} - 1 .
\label{lambda}
\end{equation}

Note that at large $\tau$ there is a
competition between the $\exp(-r/\xit)$ factor in Eq.~(\ref{sum}),
which decreases with increasing
$r$ and so would prefer to have $r$ small,
and the $\exp(-\epsilon_n \tau)$ factor which wants to have the smallest
$\epsilon_n$, which may be on a site far away.

The average value is easy to work out, since each term in Eq.~(\ref{sum})
can be evaluated separately, with the result
\begin{equation}
[S(\tau)]_{\rm av} = { \Gamma(1+1/z) \over (\epsilon_0 \tau)^{1/z} }  ,
\label{average}
\end{equation}
for $\tau \gg 1$, 
where we used Eq.~(\ref{lambda}).

However, the average is very different from the typical behavior. To see this,
we will
determine the full distribution of $S(\tau)$.
The major simplification in the calculation is that, 
at large times, the exponent in Eq.~(\ref{sum})
varies over a large range, so
the sum will be dominated by the largest single term.
Writing Eq.~(\ref{sum}) as 
\begin{equation}
S(\tau) =  A \sum_{n=0}^\infty \exp\left(-{x_n \over \xit} \right),
\end{equation}
where
\begin{equation}
x_n = n + \xit \epsilon_0 \tau e_n ,
\end{equation}
then $S(\tau)$ is given for large $\tau$ by
\begin{equation}
S(\tau) \approx  A \exp \left({- \xmin \over \xit } \right) ,
\label{stau_min}
\end{equation}
where $\xmin$ is the smallest of the $x_n$.

For the time being we will work in units of time where $\xit \epsilon_0 =
1$. At the end, we will put back this factor by
replacing $\tau$ by $ \xit \epsilon_0 \tau$.

The probability of $x_n$, is given by 
\begin{equation}
\pi_n(x_n) = {\lambda +1 \over \tau} \left({x_n - n \over \tau} \right)^\lambda,
\end{equation}
for $n < x_n < n + \tau $, and zero otherwise. Notice that the different $x_n$
have different distributions. 

We now wish to determine the probability that minimum value of the $x_n$
is $\xmin$. Let us assume first that it is $x_m$ which 
has the smallest value.  The probability that (i)
$x_m$ is the smallest, and (ii) its value is $\xmin$, is given by
\begin{equation}
{\pi_n(\xmin) \over Q_n(\xmin)} \prod_{n =0}^\infty Q_n(\xmin) ,
\label{prob_min}
\end{equation}
where 
\begin{equation}
Q_n(\xmin) = \int_{\xmin}^{n+\tau} \pi_n(x) \, dx .
\end{equation}
Since $\pi_n(x)$ vanishes for $x < n$, it follows that for $n > \xmin$, the
full range over which $\pi_n(x)$ is non-zero is included in the last integral,
and so
\begin{equation}
Q_n(\xmin) = 1, \qquad (n > \xmin ) .
\end{equation}
As a result, the product in Eq.~(\ref{prob_min})
need only be taken up to the last
integer below $\xmin$.
For $n < \xmin$, one has
\begin{equation}
Q_n(\xmin) = 1 - \left({\xmin-n \over \tau}\right)^{\lambda+1} .
\end{equation}

We will see that $\xmin \ll \tau$ and the important values of $m$ are those
where $1 \ll m \ll \tau$, and so we can rewrite Eq.~(\ref{prob_min}) as
\begin{eqnarray}
& & \qquad {\lambda + 1 \over \tau} \left({\xmin-m \over \tau} \right)^\lambda 
\nonumber \\ 
& & \qquad \times \exp \left\{ \int_0^{\xmin}
\ln \left[ 1 - \left({\xmin-x \over \tau}\right)^{\lambda+1} \right]
\, dx \right\} .
\end{eqnarray}
Expanding the log using $\xmin-x \ll \tau$ and performing the integral
gives 
\begin{equation}
{\lambda + 1 \over \tau} \left({\xmin-m \over \tau}\right)^\lambda 
\exp\left[ - { \xmin^{\lambda+2} \over (\lambda+2) \tau^{\lambda+1} } \right] .
\end{equation}
Remember that 
this is the probability that $x_m$ is the smallest of the $x_n$ and
its value if $\xmin$.
We therefore next
sum over all $m$ less than $\xmin$ to get the total probability
that the minimum is $\xmin$, i.e.
\begin{equation}
P_{min}(\xmin) = \left({\xmin \over \tau}\right)^{\lambda+1} \exp\left[
-{  \xmin^{\lambda+2} 
\over (\lambda +2) \tau^{\lambda+1} } \right] .
\end{equation}

Replacing $\tau$ by $\epsilon_0 \xit \tau$, and changing variables from $\xmin$
to $y\equiv -\ln(S(\tau) / A) $ using Eq.~(\ref{stau_min}), gives
\begin{equation}
P_y(y) = \xit\left({y \over \e0\tau}\right)^{\lambda+1} \exp\left[
-{\xit \over \lambda + 1} {y^{\lambda+2} \over (\e0 \tau)^{\lambda+1} }
\right] .
\label{py}
\end{equation}
It is easy to check that this yields the average value in Eq.~(\ref{average}).

Eq.~(\ref{py}) can be cast in a scaling form if we define
\begin{equation}
x = {y \over \tau^{1/\lambda}} =  - {\ln (S(\tau) / A) \over \tau^{1/\lambda}} .
\label{x_scale_2}
\end{equation}
Note, however, that the scaling limit involves
taking $-\ln S(\tau) \to \infty$ so the
$\ln A$ term in Eq.~(\ref{x_scale_2}) 
represents an additive {\em correction} to scaling
and can be omitted. This then leads
to Eq.~(\ref{x_scale}), in which $\lambda$ has
been replaced by $z$ using Eq,~(\ref{lambda}).
The probability
of $x$ is then a function just of $x$ (apart from a scale factor)
i.e. 
\begin{equation}
P(x) = c (cx)^{1/z} \exp\left[ -{z\over z+1} (cx)^{(z+1)/z} \right],
\end{equation}
where we have again expressed the result in terms of $z$ rather than
$\lambda$ using Eq.~(\ref{lambda}), and 
\begin{equation}
c = \left(\xit^z \e0 \right)^{1/(z+1)} .
\label{px}
\end{equation}
Eq.~(\ref{px}) is precisely Eq.~(\ref{scale_dist}) of the text.


\begin{references}

\bibitem{th}
M.~J.~Thill and D.~A.~Huse, Physica A, {\bf 15}, 321 (1995).

\bibitem{ry}
A.~P.~Young and H.~Rieger, Phys. Rev. B {\bf 53}, 8486 (1996).

\bibitem{gbh}
M.~Guo, R.~N.~Bhatt and D.~A.~Huse, 

\bibitem{dsf}
D.~S.~Fisher, Phys. Rev. Lett. {\bf 69}, 534 (1992); Phys. Rev. B {\bf
51}, 6411 (1995).

\bibitem{mccoy}
B.~M.~McCoy, Phys. Rev. Lett. {\bf 23}, 383 (1969); Phys. Rev. {\bf
188}, 1014 (1969).

\bibitem{griffiths}
R.~B.~Griffiths, Phys. Rev. Lett. {\bf 23}. 17 (1969).

\bibitem{essen}
A.~B.~Harris, Phys. Rev. B {\bf 12}, 203 (1975).

\bibitem{mw}
B.~M.~McCoy and T.~T.~Wu, Phys. Rev. B {\bf 176}, 631 (1968); {\bf 188},
982 (1969).

\bibitem{sm}
R.~Shankar and G.~Murphy, Phys. Rev.  {\bf 36}, 536 (1987).

\bibitem{yr}
A.~P.~Young and H.~Rieger, Phys. Rev. , {\bf 53}, 8486 (1996) (referred to as
YR).

\bibitem{lsm}
E.~Lieb, T. Schultz and D.~Mattis, Ann. Phys. (NY) {\bf 16}, 407 (1961).

\bibitem{pfeuty}
P.~Pfeuty, Ann. Phys. (NY) {\bf 27}, 79 (1970); Th\`ese, Universit\'e de
Paris, (1970).

\bibitem{katsura}
S.~Katsura, Phys. Rev. {\bf 127}, 1508 (1962).

\bibitem{dsf:pc}
D.~S.~Fisher (private communication).

\bibitem{ri}
H.~Rieger and F.~Igloi, cond-mat/9704152.

\bibitem{footnote2}
Note that the Hamiltonian with periodic boundary conditions, Eq.~(35) of YR,
has an extra term from the bond which connects sites $L$ and 1. 
The sign of this term depends on whether there are an even or an odd
number of excited fermions. The parity of the number of fermions is a good
quantum number, and so is well defined for each state. At zero temperature, we
only need the ground state so the fermion approach can be used. However, at
finite temperature we need to include excited states, some of which have an
even number and some an odd number of fermions. Since these
two classes of states are determined
by {\em different} Hamiltonians, the fermion approach cannot be
used at finite temperature with periodic boundary conditions. It also 
cannot be used for dynamical properties, even at $T=0$, because again
different states occur for which the fermion number has
a different parity and hence are given by a different Hamiltonian.

\bibitem{footnote1}
Note that $\epsilon_\mu$, as defined here, corresponds to $2 \epsilon_\mu$ in
YR. With our present definition, $\epsilon_\mu$ is the energy of a Fermi
excitation, see Eq.~(\protect\ref{hamdiag}),
which seems to be the most natural notation.

\bibitem{sy}
S.~Sachdev and A.~P.~Young, Phys. Rev. Lett. {\bf 78}, 2220  (1997).

\bibitem{snm}
J.~Stolze, A.~N\"oppert and G. M\"uller, Phys. Rev.  {\bf 52},
4319 (1995); H.~Asakawa, Physica A {\bf 233}, 39 (1996).

\bibitem{footnote3}
The fermion method also gives real time correlation functions, see e.g.
Ref.~\protect\cite{sy}.

\bibitem{footnote4}
Note from Eq.~(\protect\ref{AB})
that $\left(A_mB_m\right)^2 = \exp(-2i\pi c^\dagger_i c_i) = 1$.

\bibitem{mwbook}
B.~M.~McCoy and T.~T.~Wu, {\em The Two Dimensional Ising Model}, Harvard
University Press, (Cambridge, Massachusetts) (1973).

\bibitem{footnote5}
We only have limited good data for $\chi^{\rm loc}_{\rm av}$, because for most
of the values of $h_0$, the discrete set of $\tau$ values used was too coarse
to accurately perform the integral in Eq.~(\protect\ref{integral}).

\bibitem{mpaf:pc}
M.~P.A.~Fisher (private communication).

\bibitem{fy}
D.~S.~Fisher and A.~P.~Young (unpublished).

\end{references}
\end{document}